\documentclass[review]{elsarticle}

\usepackage{lineno,hyperref}

\journal{Journal of \LaTeX\ Templates}
\usepackage{amsmath, amssymb}
\usepackage{xcolor, float, caption, listings}
\usepackage{booktabs}

\begin{document}
\begin{frontmatter}

\title{A new avenue for Bayesian inference with INLA}

\author{Janet Van Niekerk\corref{mycorrespondingauthor}}
    \ead{janet.vanniekerk@kaust.edu.sa}
\author{Elias Krainski}
\author{Denis Rustand}
\author{H{\aa}vard Rue}  
\cortext[mycorrespondingauthor]{Corresponding author}

  \address{Statistics Program, Computer, Electrical and Mathematical
	Sciences and Engineering Division\\
	King Abdullah University of Science and Technology (KAUST)\\
	Thuwal 23955-6900, Kingdom of Saudi Arabia}


\begin{abstract}
    Integrated Nested Laplace Approximations (INLA) has been a
    successful approximate Bayesian inference framework since its
    proposal by \cite{rue2009}. The increased computational efficiency
    and accuracy when compared with sampling-based methods for
    Bayesian inference like MCMC methods, are some contributors to its
    success. Ongoing research in the INLA methodology and
    implementation thereof in the \textit{R} package \textit{R-INLA},
    ensures continued relevance for practitioners and improved
    performance and applicability of INLA. The era of big data and
    some recent research developments, presents an opportunity to
    reformulate some aspects of the classic INLA formulation, to
    achieve even faster inference, improved numerical stability and
    scalability. The improvement is especially noticeable for
    data-rich models.
    
    We demonstrate the efficiency gains with various examples of
    data-rich models, like Cox's proportional hazards model, an
    item-response theory model, a spatial model including prediction,
    and a 3-dimensional model for fMRI data. 
\end{abstract}


\end{frontmatter}

\section{Introduction}

Bayesian inference for latent Gaussian models can be done using 'exact' MCMC-based
or approximate methods. Integrated Nested Laplace Approximations
(INLA) \citep{rue2009} is an approximate method that deterministically
performs inference of latent Gaussian models. Latent Gaussian models
include various common and wieldy statistical models including, among
others, generalized linear mixed models (GLMM), generalized additive
mixed models (GAMM), smoothing spline models, spatial and
spatio-temporal models, survival and joint longitudinal-survival
models. The computational efficiency and accuracy of the inference
possible with INLA, act as catalysts for the uptake of the INLA
methodology in the applied sciences communities and motivates
continuous development and improvements of INLA as well as its
implementation through the \textit{R} package \textit{INLA} (R-INLA).

More than $250$ studies on COVID-19 used INLA to perform Bayesian
inference, see among others \cite{millett2020, de2020, rodriguez2020,
    kontis2020, bilal2021, davies2021, konstantinoudis2021}. Public
health studies on HIV and associated risk factors were performed using
INLA by \cite{dwyer2019, muttai2021, alene2022, tafadzwa2021}. In
environmental statistics, air pollution and its health effects were
modeled by \cite{shaddick2018, sanyal2018} while models for water and
soil pollution were considered by \cite{lillini2021}. In ecology,
\cite{pinto2020, pimont2021, lindenmayer2021} analyzed data on forest
fires and species distribution modeling was performed by
\cite{isaac2020, martinez2018, coll2019, mielke2020}. These are merely
a selected few of the recent studies that entails Bayesian inference
completed with INLA, and from this the impact of INLA on the applied
sciences is evident.

The classic formulation of the INLA methodology as presented by
\cite{rue2009} includes the linear predictors as part of the latent
field. This formulation has various benefits such as automatic
inference of the linear predictors and conditional independence in the
likelihood contributions since each datapoint depends on only one
element in the latent field. Furthermore, it gives a unified framework
to go beyond Gaussian approximations, like using simplified Laplace
approximations which are much more accurate especially for the mean.
On the downside, the latent field gets artificially large with parts that are
almost singular. As an illustration, consider a simple regression model
$y_i = a + bx_i + \epsilon_i$, with $n$ observed pairs
$(x_i, y_i)$, then the classic INLA formulation will represent this as
a graph of size $n+2$, $a$ and $b$ plus the $n$ linear predictors,
whereas the modern formulation will use a graph of size $2$ to
represent $a$ and $b$, as the linear predictors are no longer an
explicit part of the latent field. In most cases, this
does not cause a serious issue, but can be leveraged to show an advantage in
comparisons \citep{art700,art701}. But as applications are pushing the
boundaries more and more, we needed to reformulate the internal model
representation and how computations and approximations are done. Of
vital importance here, is the new Variational Bayes correction for the
mean as developed in \cite{van2021vb}, that actually allows us to
gain higher accuracy in the mean, with essentially no added cost
in the new model representation. The new model formulation adds a
clear computational gain in data-rich scenarios, where the data is
much larger than the model itself and sometimes of higher dimension.
Additionally, the new formulation gives improved numerical stability 
and a better framework for future
developments.

We will now present some basic concepts of INLA and the classic
formulation. Then in Section \ref{sec:newinla} present the modern
formulation.

\subsection{Latent Gaussian model}

Suppose we have response data $\pmb{y}_{n\times 1}$ with density
function $\pi(y|\pmb{\mathcal{X}}, \pmb\theta)$ and link function
$h(.)$, that is linked to some covariates $\pmb{Z}=\{\pmb X, \pmb U\}$
through linear predictors
\begin{equation}
    \pmb\eta_{n\times 1} = \beta_0\pmb{1} +\pmb{\beta}\pmb X  + \sum_{k=1}^K f^k(\pmb u_k)
    \label{linpred}
\end{equation}
such that $\{\pmb f\}$ are unknown functions of $\pmb U$ and
$\pmb \beta$ is the coefficients for the linear effects of $\pmb X$ on
$\pmb\eta$. The inferential aim is to estimate
$\pmb{\mathcal{X}}_{m\times 1} = \{\beta_0, \pmb\beta, \pmb f\}$,
which is the set of unobserved latent variables, or the latent field,
and $\pmb\theta$, the $p$-variate vector of hyperparameters. In the case where a
Gaussian prior is assumed for $\pmb{\mathcal{X}}$, the hierarchical
statistical model in \eqref{linpred} is a latent Gaussian model (LGM).

Latent Gaussian models are a broad class of models such as Gaussian
processes for time series, spatial and spatio-temporal data or
clustered random effects models for data with a grouping structure.
The formulation given in \eqref{linpred} can be seen as any model
where each one of the $f^{k}(.)$ terms can be written in matrix form
as $\pmb{\mathcal{A}}_k\pmb{u}_k$. But it is still not exhaustive as
volatility models can also be fitted with INLA, as in
\cite{martino2011volat} and \cite{bermudez2021volat}. 

The latent Gaussian model is the hierarchical Bayesian model defined
as follows,
\begin{eqnarray}
  \pmb y |\pmb{\mathcal{X}}, \pmb\theta_1 &\sim& \prod_{i=1}^n\pi(y_i|\mathcal{X}_i, 
                                                 \pmb\theta_1)\notag\\
  \mathcal{X}|\pmb\theta_2 &\sim& N(\pmb 0, \pmb Q_\pi^{-1}(\pmb\theta_2))\notag\\
  \pmb\theta = \{\pmb\theta_1, \pmb\theta_2\}&\sim& \pi(\pmb\theta),\label{eq:lgm}
\end{eqnarray}
where $\pmb\theta$ contains the hyperparameters from the likelihood
(like shape or precision parameters) and the latent prior (like
precision or weight parameters), and $\pmb\theta$ is allowed to assume
any (reasonable) prior. For inference we then need the following
marginal posteriors,
\begin{equation*}
    \pi(\theta_j|\pmb y) \quad \text{and}\quad 
    \pi(\mathcal{X}_j|\pmb y) 
\end{equation*}

\subsection{Classic model formulation of INLA}

We will now review the classic model formulation of \cite{rue2009}.
The linear predictors, $\pmb{\eta}$, are now included in an augmented
latent field
\begin{equation*}
    \pmb{\mathcal{X}} = \{\pmb\eta, \beta_0, \pmb\beta, \pmb f\},
\end{equation*}
that includes the linear predictors as the first $n$ elements of
$\pmb{\mathcal{X}}$ and the $m$ model components as defined in
\eqref{linpred}. This augmentation results in a singular covariance
matrix of $\pmb{\mathcal{X}}$ since $\pmb\eta$ is a deterministic
function of the other elements of $\pmb{\mathcal{X}}$. To circumvent
the singularities, we redefine
\begin{equation}
    \pmb\eta = \beta_0\pmb{1} +\pmb{\beta}\pmb X  + \sum_{k=1}^K f^k(\pmb u_k) + \pmb\epsilon,
\end{equation}
where $\pmb\epsilon\sim N(\pmb{0}, \tau^{-1}\pmb{I})$ with large fixed
$\tau$. We thus include a tiny noise term to the linear predictors to
move from a singular covariance matrix of $\pmb{\mathcal{X}}$ to a
non-singular covariance matrix. With this formulation, each datapoint
depends on only one element
of the latent field.\\ \\
Now we can write the joint density of the latent field,
hyperparameters and the data as
\begin{eqnarray}
  \pi(\pmb{\mathcal{X}},\pmb\theta| \pmb y) \propto\pi(\pmb\theta)\pi(\pmb{\mathcal{X}}|\pmb\theta)
  \prod_{i=1}^n\pi(y_i|\mathcal{X}_i,\pmb\theta)\label{eq:joint_old}
\end{eqnarray}
From here the posterior of the hyperparameters can be approximated as
\begin{eqnarray}
  \tilde{\pi}(\pmb\theta|\pmb y) &\propto
  &
    \left.
    \frac{\pi(\pmb{\mathcal{X}},\pmb\theta| \pmb y) }
    {\pi_{G}(\pmb{\mathcal{X}}|\pmb\theta, \pmb y) }\right |_{\pmb{\mathcal{X}} = \pmb\mu(\pmb\theta)},\label{eq:hyperpost}
\end{eqnarray}
where $\pi_{G}(\pmb{\mathcal{X}}|\pmb\theta, \pmb y)$ is a Gaussian
approximation to the true $\pi(\pmb{\mathcal{X}}|\pmb\theta, \pmb y) $
using the Laplace method \citep{tierney1989fully} by matching the mode
and the curvature around the mode of
$\pi(\pmb{\mathcal{X}}|\pmb\theta, \pmb y) $. Next the posterior
marginals can be approximated by
\begin{eqnarray}
  \tilde{\pi}(\mathcal{X}_j|\pmb y) &=& \int \tilde{\pi}(\mathcal{X}_j|\pmb{\theta},\pmb y)
                                        \tilde{\pi}(\pmb\theta|\pmb y)d\pmb\theta\notag\\
  \tilde{\pi}(\theta_j|\pmb y) &=& \int \tilde{\pi}(\pmb\theta|\pmb y)d\pmb\theta_{-j}.\notag
\end{eqnarray}
The approximation $\tilde{\pi}(\mathcal{X}_j|\pmb{\theta},\pmb y)$ can
be obtained in two ways, either from the Gaussian approximation
$\pi_{G}(\pmb{\mathcal{X}}|\pmb\theta, \pmb y)$ in
\eqref{eq:hyperpost}, or from another Laplace approximation step as
follows:
\begin{equation}
    \tilde{\pi}(\mathcal{X}_j|\pmb{\theta},\pmb y)
    \propto \left. \frac{\pi(\pmb{\mathcal{X}},\pmb\theta| \pmb y)}{
          \pi_{G}(\pmb{\mathcal{X}}_{-j}|\mathcal{X}_j,\pmb\theta,
          \pmb y)}\right|_{
        \pmb{\mathcal{X}}_{-j} = \pmb\mu_{-j}(\pmb\theta)},\label{eq:fulllaplace}
\end{equation}
where $\pmb\mu_{-j}(\pmb\theta)$ is the mode of
$\pi_{G}(\pmb{\mathcal{X}}_{-j}|\mathcal{X}_j,\pmb\theta, \pmb y)$.
The Gaussian strategy is computationally cost effective but in the
case of non-Gaussian likelihoods can be quite inaccurate, whereas the
extra Laplace approximation step (the nested one) is computationally
more expensive but provides sufficiently accurate inference. The
unified treatment of model components and the linear predictors, allow
us to compute the simplified Laplace approximation (see \cite{rue2009} for details), say, for all
components, in a unified way.

\subsection{Proposal}

In many cases, the augmentation of the latent field by the linear
predictors does not increase the computational cost of the inference
by much, but for low-dimensional models with high-dimensional data,
the nested Laplace approximation gets more computationally expensive, mainly
due to this augmentation. As noted, initially the linear predictors
were included in the latent field definition to perform inference in a
unified way. The other options at the time were to not include them
and then calculate their posteriors, post-inference of the latent
field. In the case of the posterior Gaussian assumption for the latent
field and linear predictors this approach is cost effective but can be
quite inaccurate since the tiny errors compound in the post-inference
framework. Another option is to construct skew-normal marginals with a
Gaussian copula for the linear predictors from the inference of the
latent field. This option is computationally expensive but more
accurate and essentially negates the computational benefit of removing
the linear predictors from the latent field.

Recent developments by \cite{van2021vb} enables us to define a new
framework for approximate Bayesian inference using INLA without the
linear predictor in the latent field formulation, defined as an extended latent 
Gaussian model by \cite{art700}. The low-rank
Variational Bayes correction to the posterior means of a Gaussian
latent field proposed by \cite{van2021vb} can now be used to
efficiently correct the posterior means of the latent field,
post-inference of the latent field, and then obtain the marginal
posteriors of the linear predictors from integrating the Gaussian
conditional marginals. Without this efficient correction, the results
might be inaccurate. We present the proposed methodology in
Section~\ref{sec:meth} and illustrate this new approach with
examples in Section~\ref{sec:examples} and Section~\ref{sec:fmri}.

\section{Methodology}\label{sec:meth}

In this section we propose the modern formulation of the INLA
methodology, and some details on the implementation thereof to attain
computational efficiency and sufficiently accurate inference.

\subsection{Modern model definition and INLA}\label{sec:newinla}

The main enhancement of the classic formulation is that we do not
augment the latent field with the ``noisy" linear predictors. The
latent field of dimension $m$ is defined as
\begin{equation*}
    \pmb{\mathcal{X}}= \{\beta_0, \pmb\beta, \pmb f\},
\end{equation*}
with Gaussian prior
$\pmb{\mathcal{X}}|\pmb\theta\sim N(\pmb 0, \pmb Q^{-1}(\pmb\theta))$,
and the $n$ linear predictors are defined as
\begin{eqnarray}
  \pmb\eta = \pmb A\pmb{\mathcal{X}},
\end{eqnarray}
with $\pmb A$ a sparse design matrix that links the linear predictors
to the latent field.

From this formulation the joint density of the latent field,
hyperparameters and the data is derived as
\begin{equation}
    \pi(\pmb{\mathcal{X}},\pmb\theta| \pmb y) \propto \pi(\pmb\theta)\pi(\pmb{\mathcal{X}}|\pmb\theta)
    \prod_{i=1}^n\pi(y_i|(\pmb A\pmb{\mathcal{X}})_i,\pmb\theta).\label{eq:joint_new}
\end{equation}
With this formulation the first step is to find the mode and the
Hessian at the mode of $\tilde{\pi}(\pmb\theta|\pmb y)$ defined as
follows:
\begin{equation}
    \tilde{\pi}(\pmb\theta|\pmb y) = \left.
      \frac{\pi(\pmb{\mathcal{X}},\pmb\theta| \pmb y)}{
          \pi_G(\pmb{\mathcal{X}}|\pmb\theta, \pmb y)}\right |_{\pmb{\mathcal{X}}=\pmb\mu(\pmb\theta)}.
    \label{eq:pihyper}
\end{equation}  
The Gaussian approximation
$\pi_G(\pmb{\mathcal{X}}|\pmb\theta, \pmb y)$ to
$\pi(\pmb{\mathcal{X}}|\pmb\theta, \pmb y)$ is calculated from a
second order expansion of the likelihood around the mode of
$\pi(\pmb{\mathcal{X}}|\pmb\theta, \pmb y)$, $\pmb\mu(\pmb\theta)$ as
follows
\begin{eqnarray*}
  \log\left(\pi(\pmb{\mathcal{X}}|\pmb\theta, \pmb y)\right)
  &\propto& -\frac{1}{2}\pmb{\mathcal{X}}^\top\pmb
            Q(\pmb\theta)\pmb{\mathcal{X}} +
            \sum_{i=1}^n\left(b_i(\pmb A\pmb{\mathcal{X}})_i-\frac{1}{2}c_i(\pmb A\pmb{\mathcal{X}})^2_i \right)\\
  &=&  -\frac{1}{2}\pmb{\mathcal{X}}^\top\left(\pmb Q(\pmb\theta) +
      \pmb A^\top\pmb D \pmb A\right)\pmb{\mathcal{X}} - \pmb b^\top\pmb A\pmb{\mathcal{X}}
\end{eqnarray*}
where $\pmb b$ is an $n$-dimensional vector with entries $\{b_i\}$ and
$\pmb D$ is a diagonal matrix with $n$ entries $\{c_i\}$. Note that
both $\pmb b$ and $\pmb D$ depend on $\pmb\theta$, so the Gaussian
approximation is for a fixed $\pmb\theta$. The process is iterated to
find $\pmb b$ and $\pmb D$ that gives the Gaussian approximation
at the mode, $\pmb\mu(\pmb\theta)$, so that
\begin{equation}
    \pmb{\mathcal{X}}|\pmb\theta, \pmb y \sim
    N\left(\pmb\mu(\pmb\theta),
      \pmb Q^{-1}_{\mathcal{X}}(\pmb\theta)\right).\label{eq:xmulticond}
\end{equation}
Note that with the modern formulation, the graph of the Gaussian
approximation consists of two components,
\begin{enumerate}
\item $\mathcal{G}_p$: the graph obtained from the prior of the latent
    field through $\pmb Q(\pmb\theta)$
\item $\mathcal{G}_d$: the graph obtained from the data based on the
    non-zero entries of $\pmb A^\top \pmb A$
\end{enumerate}
Next, the marginal conditional posteriors of the elements of
$\pmb{\mathcal{X}}$ is calculated from the joint Gaussian
approximation in \eqref{eq:pihyper} as
\begin{equation}
    \mathcal{X}_j|\pmb\theta,\pmb y \sim
    N\left(\left(\pmb\mu(\pmb\theta)\right)_j,
      \left(\pmb Q^{-1}_{\mathcal{X}}(\pmb\theta)\right)_{jj}\right).
    \label{eq:margcondxj}
\end{equation}
To get the marginal posterior of $\mathcal{X}_j$ we need to
(numerically) integrate $\pmb\theta$ out from \eqref{eq:margcondxj}
using $K$ integration points $\pmb\theta_k$ and area weights
$\delta_k$ defined by some numerical integration scheme
\begin{equation}
    \tilde{\pi}(\mathcal{X}_j| \pmb y) = \int
    \pi_G(\mathcal{X}_j|\pmb\theta, \pmb y)d\pmb\theta
    \approx \sum_{k=1}^K\pi_G(\mathcal{X}_j|\pmb\theta_k, \pmb y)
    \tilde{\pi}(\pmb\theta_k|\pmb y)\delta_k.\label{eq:marginalx}
\end{equation}
Even though we propose a Gaussian for the conditional marginal
posterior of $\mathcal{X}_j$ in \eqref{eq:margcondxj}, the marginal
posterior of $\mathcal{X}_j$ will be skewed due to the integration as
in \eqref{eq:marginalx}.

\subsection{Marginal posteriors of the linear predictors
    $\tilde\pi(\eta_i|\pmb y)$}
In the classic formulation, the marginal posteriors of the linear
predictors are automatically calculated since they constitute the
first $n$ elements of the latent field. With the new formulation as
set out in Section \ref{sec:newinla}, we need to calculate the
marginal posteriors of the linear predictors after we calculated
$\tilde\pi(\mathcal{X}_j| \pmb y)$ and $\tilde\pi(\pmb\theta|\pmb y)$,
and we need to do this in an efficient way if we want computational
benefit compared to the classic INLA framework.

In order to calculate $\tilde\pi(\eta_i|\pmb y)$, we first calculate
$\tilde\pi(\eta_i|\pmb\theta,\pmb y)$. We postulate a Gaussian density
for $\eta_i|\pmb\theta,\pmb y$ such that
$\tilde\pi(\eta_i|\pmb\theta,\pmb y) = \pi_G(\eta_i|\pmb\theta,\pmb
y)$, with mean
\begin{equation*}
    E(\pmb\eta|\pmb\theta,\pmb y) = \pmb A E(\pmb{\mathcal{X}}|\pmb\theta,\pmb y) =  \pmb A\pmb\mu(\pmb\theta)
\end{equation*}
and covariance matrix
\begin{equation}
    \text{Cov}(\pmb\eta|\pmb\theta,\pmb y) = \pmb A \text{Cov}(\pmb{\mathcal{X}}|\pmb\theta,\pmb y)\pmb A^\top,
    \label{eq:covlinpred}
\end{equation}
from \eqref{eq:margcondxj}, and since
\begin{equation*}
    \pmb\eta = \pmb A\pmb{\mathcal{X}}.
\end{equation*}
Note that $\pmb A$ is a sparse matrix so the computations are
efficient. However, in the Gaussian approximation to
$\pi(\pmb{\mathcal{X}}|\pmb\theta,\pmb y)$ we calculate the precision
matrix $\pmb Q_{\mathcal{X}}$, but to use \eqref{eq:covlinpred} we
need the covariance matrix $\pmb Q^{-1}_{\mathcal{X}}$, which is very
expensive to calculate and store. Conversely, for the marginal conditional
posterior of $\eta_i$, we only need the $i^\text{th}$ diagonal element
of
$\pmb A \text{Cov}(\pmb{\mathcal{X}}|\pmb\theta,\pmb y)\pmb A^\top$.

Define the sparse selected inverse of a symmetric sparse matrix
$\pmb Q$ with graph $\mathcal{G}_\mathcal{X}$ as the symmetric matrix
$\pmb C$ with elements
$\{C_{il} : i\sim_{\mathcal{G}_\mathcal{X}}l \text{ or } i=l\}$. The
elements of the sparse selected inverse is computed from the Cholesky
decomposition of $\pmb Q = \pmb L \pmb L^\top$ as follows (see
\cite{takahashi1973} and \cite{rue2009} for details)
\begin{equation*}
    C_{ij} = \frac{\delta_{ij}}{L_{ii}^2} - \frac{1}{L_{ii}}\sum_{\substack{k>i\\L_{ki}\neq 0}}L_{ki}C_{kj},
\end{equation*}
where $\delta_{ij} = 1$ for $i=j$ and zero otherwise. This way we only
calculate the necessary elements of the inverse and using only the
non-zero elements of $\pmb L$ provided $\pmb Q$ has been properly
reordered. Now let $\pmb C$ be a sparse selected inverse of
$\pmb Q_\mathcal{X} = \pmb Q + \pmb A^\top\pmb D\pmb A$ based on the
graph $\mathcal{G}_\mathcal{X} = \{\mathcal{G}_p,\mathcal{G}_d\}$,
then
\begin{equation}
    \text{Var}(\eta_j|\pmb\theta,\pmb y) = (\pmb A\pmb C\pmb
    A^\top)_{jj}
    =\sum_{il}\pmb A_{ji}C_{il}(\pmb A^\top)_{lj} = \sum_{il}A_{ji}A_{jl}C_{il}.
    \label{eq:varlinpred}
\end{equation}
Note that in $\pmb Q_\mathcal{X}$ we get elements from
$\pmb A^\top\pmb D \pmb A$ whereas for the variance of the linear
predictors we need the diagonal of $\pmb A \pmb C \pmb A^\top$. Since
$\pmb Q_\mathcal{X}$ is based on graph $\mathcal{G}_\mathcal{X}$ we
know that for $i\sim_{\mathcal{G}_\mathcal{X}}j$, we have
\begin{equation*}
    \left(\pmb Q_\mathcal{X}\right)_{ij} = \sum_l (\pmb A^\top)_{ij}D_{il}A_{jl}  = \sum_l A_{ji}D_{il}A_{jl} 
\end{equation*}
with $A_{ji}A_{jl}\neq 0 $ for at least one $j$ since
$i\sim_{\mathcal{G}_\mathcal{X}}l$ based on $\mathcal{G}_\mathcal{X}$.
Also, in \eqref{eq:varlinpred} we then know that if
$i\sim_{\mathcal{G}_\mathcal{X}}l$ then we must have
$A_{ji}A_{jl}\neq 0$. This ensures a sensible calculation of the
variances defined in \eqref{eq:varlinpred}.

Now that we can calculate the mean and the variance of the conditional
posterior of $\eta_j$ we can calculate the marginal posterior of
$\eta_j$ using \eqref{eq:marginalx} as follows:
\begin{eqnarray}
  \eta_j|\pmb\theta,\pmb y &\sim& N(\mu_j(\pmb\theta), \sigma^2_j(\pmb\theta))\notag\\
  \mu_j(\pmb\theta) &=& (\pmb A\pmb\mu(\pmb\theta))_j\notag\\
  \sigma^2_j(\pmb\theta) &=& \sum_{il}A_{ji}A_{jl}C_{il}\notag\\
  \tilde{\pi}(\eta_j| \pmb y)
                           &\approx
                                & \sum_{k=1}^K\pi_G(\mathcal{\eta}_j|
                                  \pmb\theta_k, \pmb y)\tilde{\pi}(\pmb\theta_k|\pmb y)\delta_k,\label{eq:marginaleta}
\end{eqnarray}
so that $\text{E}(\eta_j|\pmb y) = \mu_j$ and
$\text{Var}(\eta_j|\pmb y) = \sigma_j^2$.

\subsection{Low-rank correction using Variational Bayes for increased
    accuracy}\label{sec:VB}

The posterior means of $\pmb\eta$ and $\pmb{\mathcal{X}}$ might be inaccurate based on
the Gaussian assumption of the conditional posterior, especially in
the case of a non-Gaussian likelihood. The recent proposed Variational
Bayes correction to Gaussian means by \cite{van2021vb}, can be used
to efficiently calculate an improved mean for the marginal posteriors
of the linear predictors, by improving the posterior means of the
latent field.\\ \\
\cite{zellner1988optimal} showed that Bayes' rule is a $100\%$ efficient information processing rule. Based on this work, we can solve for the posterior based on the following variational function,
\begin{equation*}
E_{q(\pmb{\mathcal{X}}|\pmb{y})}\left[-\log l(\pmb{\mathcal{X}}|\pmb{y})\right] +
\text{KLD}\left[q(\pmb{\mathcal{X}}|\pmb{y})||\pi(\pmb{\mathcal{X}})\right]
\end{equation*}
where $q(.)$ is a member of the variational class, $\pi(.)$ is the prior and $l(.)$ is the likelihood function. Instead of using this method to calculate the posterior of $\pmb{\mathcal{X}}|\pmb\theta$, we rather use the variational framework to calculate a posterior correction to the mean from the Gaussian approximation at the mode as in \eqref{eq:xmulticond} and \eqref{eq:margcondxj}. There are other ways to find element-wise corrections to the elements of the Gaussian mean, but with element-wise changes there is no guarantee that the overall effects would be an improvement. With this variational formulation on the joint mean, we are sure of a joint improvement. Also, this formulation enables a low-rank correction where influential elements are corrected and the effect of these corrections are then propagated to the other elements such that the corrections imply an improved joint mean. We present the details of this approach next.\\ \\
From Section \ref{sec:newinla}, \eqref{eq:xmulticond} a mean,
$\pmb\mu(\pmb\theta)$, of $\pmb{\mathcal{X}}|\pmb y, \pmb\theta$ is
calculated in the Gaussian approximation at the mode of
$\pi(\pmb{\mathcal{X}}|\pmb y, \pmb\theta)$. Using a Variational
framework we can improve on this mean as proposed by
\cite{van2021vb}. To this end, we formulate a low-rank variational
Bayes correction of size $p<m$ to $\pmb\mu(\pmb\theta)$, such that the
improved mean is
\begin{equation}
    \pmb\mu^*(\pmb\theta)=\pmb\mu(\pmb\theta)+\pmb M\pmb\lambda,
\end{equation}
where $\pmb M$ is a matrix that propagates the correction made to $p$
nodes to the rest of the latent field. The propagation matrix $\pmb M$
is derived from selected columns of
$\pmb Q_{\mathcal{X}}^{-1}(\pmb\theta)$ (corresponding to the $p$
nodes of explicit correction) as presented by \cite{van2021vb}. The
explicit corrections $\pmb\lambda$, are solved for iteratively using
the variational function
\begin{equation*}
    \arg_{\pmb\lambda} \min \left(E_{\mathcal{X}|\pmb y, \pmb\theta\sim N(\pmb\mu(\pmb\theta)
          + \pmb M\pmb\lambda, \pmb Q^{-1}_{\mathcal{X}}(\pmb\theta))} \left[-\log l(\pmb{\mathcal{X}}|\pmb{y})\right]
      +  \frac{1}{2}\left(\pmb\mu(\pmb\theta) +
        \pmb M\pmb\lambda\right)^\top\pmb Q(\pmb\theta)\left(\pmb\mu(\pmb\theta)+\pmb M\pmb\lambda\right)\right),
\end{equation*}
where
\begin{equation*}
    \log l(\pmb{\mathcal{X}}|\pmb{y}) = \sum_{i=1}^n \log\pi(y_i|(\pmb A\pmb{\mathcal{X}})_i)
\end{equation*} 
The expected log-likelihood is calculated using
Gauss-Hermite quadrature with $n_J$ nodes $\pmb x$ and $n_J$
weights $\pmb w$, as follows:
\begin{equation*}
    E_{\mathcal{X}|\pmb y, \pmb\theta\sim N(\pmb\mu(\pmb\theta)
        + \pmb M\pmb\lambda, \pmb Q^{-1}_{\mathcal{X}}(\pmb\theta))}
    \left[-\log l(\pmb{\mathcal{X}}|\pmb{y})\right] = \sum_{i=1}^n
    \sum_{j=1}^{n_J}w_j\log\pi[y_i|(\sigma_i(\pmb\theta)x_j + \mu_i(\pmb\theta))],
\end{equation*}
where
$\pmb Q_{\mathcal{X}}(\pmb\theta) =\pmb L(\pmb\theta)\pmb L^\top
(\pmb\theta)$ from a Cholesky decomposition, $\sigma_i(\pmb\theta)$ is from \eqref{eq:marginaleta} and 
$\mu_i(\pmb\theta)=(\pmb A\pmb\mu^*(\pmb\theta))_i$. Then a second-order
Taylor series expansion around $\pmb\lambda = \pmb 0$ is calculated
and used to approximate
$E_{\mathcal{X}|\pmb y, \pmb\theta\sim N(\pmb\mu(\pmb\theta) + \pmb
    M\pmb\lambda, \pmb Q^{-1}_{\mathcal{X}}(\pmb\theta))} \left[-\log
  l(\pmb{\mathcal{X}}|\pmb{y})\right]$. This correction is
computationally efficient and achieves accuracy for the mean similar to a full
integrated nested Laplace approximation as shown by \cite{van2021vb}.
Now this improved mean can be used in \eqref{eq:marginaleta} for
posterior inference of $\pmb\eta$ as follows:
\begin{eqnarray}
  \eta_j|\pmb\theta,\pmb y &\sim& N(\mu_j(\pmb\theta), \sigma^2_j(\pmb\theta))\notag\\
  \mu_j(\pmb\theta) &=& (\pmb A\pmb\mu^*(\pmb\theta))_j\notag\\
  \tilde{\pi}(\eta_j| \pmb y)
                           &\approx&
                                     \sum_{k=1}^K\pi_G(\mathcal{\eta}_j|\pmb\theta_k, \pmb y)
                                     \tilde{\pi}(\pmb\theta_k|\pmb y)\delta_k.\label{eq:marginaleta_imp}
\end{eqnarray}

\subsection{Other developments used in INLA}

To find the mode of $\tilde{\pi}(\pmb\theta|\pmb y)$ we need to
perform unconstrained optimization. Based on the work of
\cite{abdulfattah2022} we use the smart gradient approach to more
accurately and efficiently calculate the gradient, and hence find the
mode and the Hessian at the mode more accurately.

Numerical methods for estimating the gradient such as the forward
difference, backward or central finite difference methods
traditionally use the canonical basis. Instead of using the canonical
basis $\pmb e$, we construct a new orthonormal basis using the
previous descent directions and the Modified Gram-Schmidt (MGS)
orthogonalization. Suppose $\tilde\nabla f$ is the numerical gradient
of the gradient $\nabla f$ of
$f(\pmb\theta) = \tilde{\pi}(\pmb\theta|\pmb y)$, and the dimension of
$\pmb\theta$ is $q$. Now, at iteration $k$, we use the previous $q$
descent directions and construct new directions as follows:
\begin{equation*}
	\pmb d^{(k)} = \frac{\Delta \pmb\theta^{(k)}}{|| \Delta \pmb\theta^{(k)}||}
\end{equation*}
Then we subtract those directions already included, through
projections, as to get a unique contribution for each direction. From
these projected directions we can construct an orthonormal basis using
MGS, contained in the columns of $\pmb G^{(k)}$, that spans the same
space as the original direction vectors, and the estimated gradient at
iteration $k$ can then be calculated as follows:
\begin{equation*}
	\tilde\nabla_{\pmb d} f^{(k)}(\pmb\theta^{(k)})={\pmb
	{G}^{(k)}}^{-\top}\tilde\nabla_{\pmb e}
	h^{(k)}(\pmb\psi),\quad h^{(k)}(\pmb\psi) = f(\pmb\theta^{(k)}+\pmb G^{(k)}\pmb\psi).
\end{equation*}
This smart gradient framework enables a more accurate and expedited
location of the mode of $\tilde{\pi}(\pmb\theta|\pmb y)$.\\ \\
For high performance computing environments, INLA has been parallelized using the \textit{OpenMP} and \textit{PARDISO} libraries. \textit{PARDISO} is a state-of-the-art library containing parallelized versions of efficient sparse linear algebra solvers that is employed at all stages of the INLA algorithm. Another level of parallelism is introduced for calculating $\tilde{\pi}(\pmb\theta|\pmb y)$ at multiple $\pmb\theta$, and finding the mode of $\tilde{\pi}(\pmb\theta|\pmb y)$ through a parallel line search algorithm. This multi-level parallelism has been shown to lead to speed-up factors of up to $10$ for large models. This development allows the user to efficiently employ the INLA methodology in multi-core architectures. For more details see \cite{gaedke2022}.

\section{Examples}\label{sec:examples}

In this section we illustrate the novelty of the new framework with
some well-known data-rich models, the Cox proportional hazards model
and item-response theory models. We will also illustrate that the
improved numerical stability can make a difference, like in models for
spatial prediction. We deliberately chose the simulated examples to give moderate speedup,
as due to differences in the computational scaling, we can essentially
achieve any speedup we desire. All results were computed on an Ubuntu
system with Intel Xeon Gold 6230R at 2.1GHz processors and 772GB of
RAM.

\subsection{Cox proportional hazards model}

The likelihood of Cox's proportional hazards model is equivalent to
the likelihood of a specific Poisson process as shown by
\cite{holford1980analysis} and \cite{laird1981covariance}. This
reformulation is necessary for the Cox's proportional hazards model to
be classed as a latent Gaussian model. This process entails data
augmentation since for each datapoint in the Cox model, we need
$k\leq K$ datapoints for the Poisson regression model, with $K$ the
number of bins in the baseline hazard where the baseline hazard is
assumed piecewise log-constant.
\subsubsection{Data augmentation}
Suppose there is an observed time and censoring indicator $(t, d)$ and
we assume a baseline hazard with $B$ bins, $(s_1,s_2,...,s_B)$, then
for each bin until the observed time $t$ we will have $k$ augmented
datapoints such that there are $k-1$ datapoints with a Poisson
distribution with mean
$\lambda_j = \exp (\eta_j)(s_{j+1}-s_j), j=1,...,k-1$ that is observed
to be $0$, and one datapoint observed as $0$ ($t$ is censored) or $1$
($t$ is not censored) that follows a Poisson distribution with mean
$\lambda_k = \exp( \eta_k)(t-s_k)$, where
\begin{equation}
    \eta_j = \pmb\beta \pmb X + \sum_{l=1}^Lf^l(\pmb u_l) + b_j, \quad t\in (s_{j-1},s_j]
\end{equation}
with $\exp(b_j)$ as the baseline hazard for bin $j$, and we assume a
random walk model of order one or two for $\pmb b$ (see
\cite{martino2011} for more details).

This data augmentation scheme clearly results in a much larger dataset
than the original dataset, and in this case including the linear
predictors in the latent field could result in much higher
computational cost. We illustrate the advantage of the new approach
with Cox proportional hazards models in the next section.
\subsubsection{Simulated example}
We simulate survival data for $n$ patients using the following very simple Cox
proportional hazards model
\begin{equation}
    h_i(t) = h_0(t)\exp(\beta x_i)=1.2t^{0.2}\exp\left(0.1x_i\right),\quad 1=1,2,...,n,
    \label{eq:cox_sim}
\end{equation}
where $x$ is a scaled and centered continuous covariate, and the
baseline hazard, $h_0(t)$ is estimated using a scaled random walk order one
model (see \cite{sorbye2014}) with $50$ bins. We also consider four different values of $n$
which are $n=10^{2}$, to $10^{5}$, illustrate the scalability of the
new approach.
\begin{table}[H]
    \centering
    \begin{tabular}{@{}rrrr}\toprule
      $n$ & Augmented size & classic INLA (s)& modern INLA (s)\\ 
      \cmidrule{1-4}
      $10^2$ & $1\,327$ & $1.6$ & $0.1$\\
      $10^3$ & $12\,657$ & $1.3$ & $0.4$\\
      $10^4$ & $131\,807$ & $10.2$ & $2.3$\\
      $10^5$ & $1\,302\,413$ & $113.3$ & $22.5$\\
      \bottomrule
    \end{tabular}
    \caption{Results from simulation of Cox proportional hazards model
        \eqref{eq:cox_sim}.}
    \label{tab:cox_sim}
\end{table}
In Table \ref{tab:cox_sim} the benefit of the modern formulation is
clear when comparing the time used for inference between the classic
and modern formulations. Note that when the augmented data set is large,
the original INLA framework fails to produce results due to the large
addition of linear predictors to the latent field, which is not the
case in the modern formulation. The accuracy of the inference for the
model components, $\beta$ and $h_0(t)$ is not compromised due to the
VB correction as mentioned in Section \ref{sec:VB}.

\subsubsection{Real example: AIDS}

Acquired immunodeficiency syndrome (AIDS) is a chronic, potentially
life-threatening condition caused by the human immunodeficiency virus
(HIV), and it is incurable but treatable. More than 37 million people
were infected by HIV by 2020 and it is thus a virus of global concern.

\begin{figure}
    \includegraphics[height = 6cm]{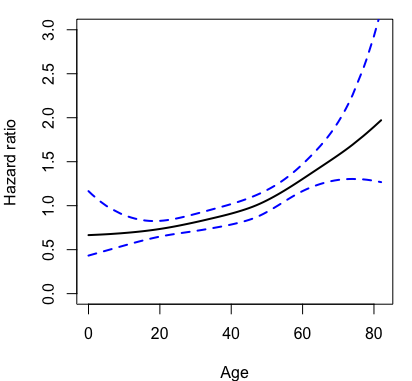}
    \hspace*{5mm}
    \includegraphics[height = 6cm]{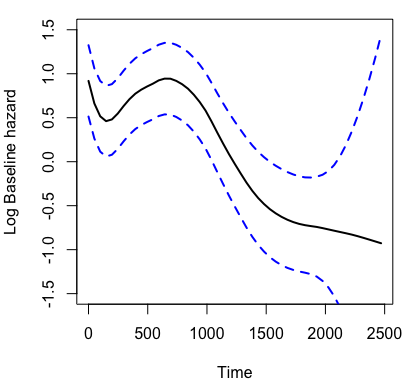}
    \caption{Inference for the non-linear effect of age and the
        baseline hazard}
    \label{fig:cox_real}
\end{figure}

This dataset contains the information of $2843$ patients residing in
Australia who were diagnosed with AIDS. The clinical aim is to
investigate how age (Age) and transmission category (TC) affects the
hazard rate of death due to AIDS-related complications, for more
details on the dataset see \cite{venables2013}. Another interest is
the effect of Zidovudine (AZT) therapy which was introduced in
Australia in July 1987. To this end we added the information on AZT
use for those cases after July 1987 to see if there is an effect on
the hazard of death. The model we consider is
\begin{equation*}
    h(t) = h_0(t)\exp\left(\beta_0 + \beta_1\text{AZT} + \pmb\beta_{\text{TC}} + f(\text{Age}) \right),
\end{equation*}
where $f(\text{Age})$ is an unknown non-linear function of age and the
log of the baseline hazard $\log h_0(t)$ are both estimated by a scaled
random walk order two prior model (see \cite{lindgren2008}) with precision parameters $\tau_f$ and $\tau_h$, respectively.\\ \\
The augmented dataset contains $24\ 830$ records and the analysis is
performed in $0.8$ seconds with the new framework. The results are
presented in Table \ref{tab:cox_real} and Figure \ref{fig:cox_real}.
We note that the use of AZT and the category TC=$3$ are significantly
associated with a decreased hazard of death. We also see that age has
a significant effect on the hazard, such that the hazard increases with age.
\begin{table}[H]
    \centering
    \begin{tabular}{@{}lrrrr@{}}\toprule
      & Posterior mean  & $0.025$-Quantile & $0.975$-Quantile & Hazard Ratio\\
      \cmidrule{2-5}
$\beta_0$     & $0.197$      &$-0.207$      &$0.603$  & \\
$\beta_1$ 		& -$0.466^*$ 		& $-0.571$	&$-0.362$ & $0.627^*$\\ 
$\beta_{TC1}$    & $-0.126$     &$-0.424$      &$ 0.173$ & $0.882$\\
$\beta_{TC2}$     & $-0.418$     &$-0.873$      &$0.038$ & $0.658$\\
$\beta_{TC3}$      & $-0.724^*$     &$-1.202$      &$-0.245$ & $0.485^*$\\
$\beta_{TC4}$    & $ 0.283$     &$-0.094$      &$ 0.66$ & $1.328$\\
$\beta_{TC5}$    & $ 0.182$     & $-0.088$      &$ 0.447$ & $1.2$\\
$\beta_{TC6}$  & $0.073$     &$-1.153$      &$ 1.283$ & $1.076$\\
$\beta_{TC7}$   & $ 0.107$     &$-0.213$      &$ 0.426$ & $1.112$\\
$\tau_f$        & $1152.17$    & $11.612$      &$7704.94$ & \\
$\tau_h$        & $1.94$       &$0.666$        &$4.63$ & \\
      \bottomrule
    \end{tabular}
    \caption{Inference for the Australian AIDS dataset.\\
    $^*$ Statistically significant effect}
    \label{tab:cox_real}
\end{table}

\subsection{IRT models}

The Item Response Theory (IRT) models are often used in the analysis
of standardized tests where multiple students answer multiple
questions. In this setting IRT models are considered to estimate the
subjects ability on some school subject. However, IRT models are also
being applied in other fields, such as in social sciences to estimate
the ideal point orientation regarding some matter, medicine for
diagnosis from patient outcomes and marketing, as one can see from
\cite{linden2016handbirt}.

There are a number of R packages to implement IRT models, considering
different parameterization, extensions and inference paradigm, see
\cite{choi2019irtpackages}. For example, the Bayesian IRT models can
be fitted using Stan, as shown in \cite{burkner2021bayesirt}. But
specialized efficient algorithms and implementation exists, see
\cite{martin2011MCMCpack}. As in many cases the IRT models can be
framed as conventional statistical models and, consequently, conventional statistical
modeling software such as INLA as illustrated in
\cite{mair2022irtinla}.

In the simplest case the data available is coded as zero or one as the
answer from subject $j$ to item $i$,
\begin{eqnarray}\label{eq1}%
     y_{ij} & \sim & \textrm{Bernoulli}(p_{ij}) \notag\\
     p_{ij} & = & 1/(1+\exp(-\beta_i(\kappa_j-\alpha_i)))\label{eq:sim_irt}
\end{eqnarray}
where $\kappa_j$ is the subject score, $\alpha_i$ is the item
difficulty and $\beta_i$ is the item discrimination. Therefore, we
have $n+2m$ parameters, where $n$ is the number of subjects and $m$ is
the number of items, with $nm$ available data.

\subsubsection{Simulated example}

We simulate data for $n$ students responding to $20$ items. For each
simulation, the model parameters were drawn randomly as
$\beta_i\sim Gamma(20,20),\ \alpha_i\sim N(0,1),\ \kappa_j\sim N(0,1)$.
We consider various values for $n$ and present the results in
Table~\ref{tab:sim_irt}.

\begin{table}[H]
    \centering
    \begin{tabular}{@{}lcc@{}}\toprule
      $n$ &  classic INLA (s) & modern INLA (s)\\ 
      \cmidrule{1-3}
      $10^{2}$ &  $9.3$ & $3.0$\\ 
      $10^{3}$  &  $29.0$ & $8.4$\\ 
      $10^{4}$ & $226.4$ & $73.8$\\ 
      \bottomrule
    \end{tabular}
    \caption{Results from simulation of IRT model \eqref{eq:sim_irt}}
    \label{tab:sim_irt}
\end{table}

\subsubsection{Real example: ENEM}

\begin{figure}[h] \centering
    \includegraphics[width=0.99\textwidth]{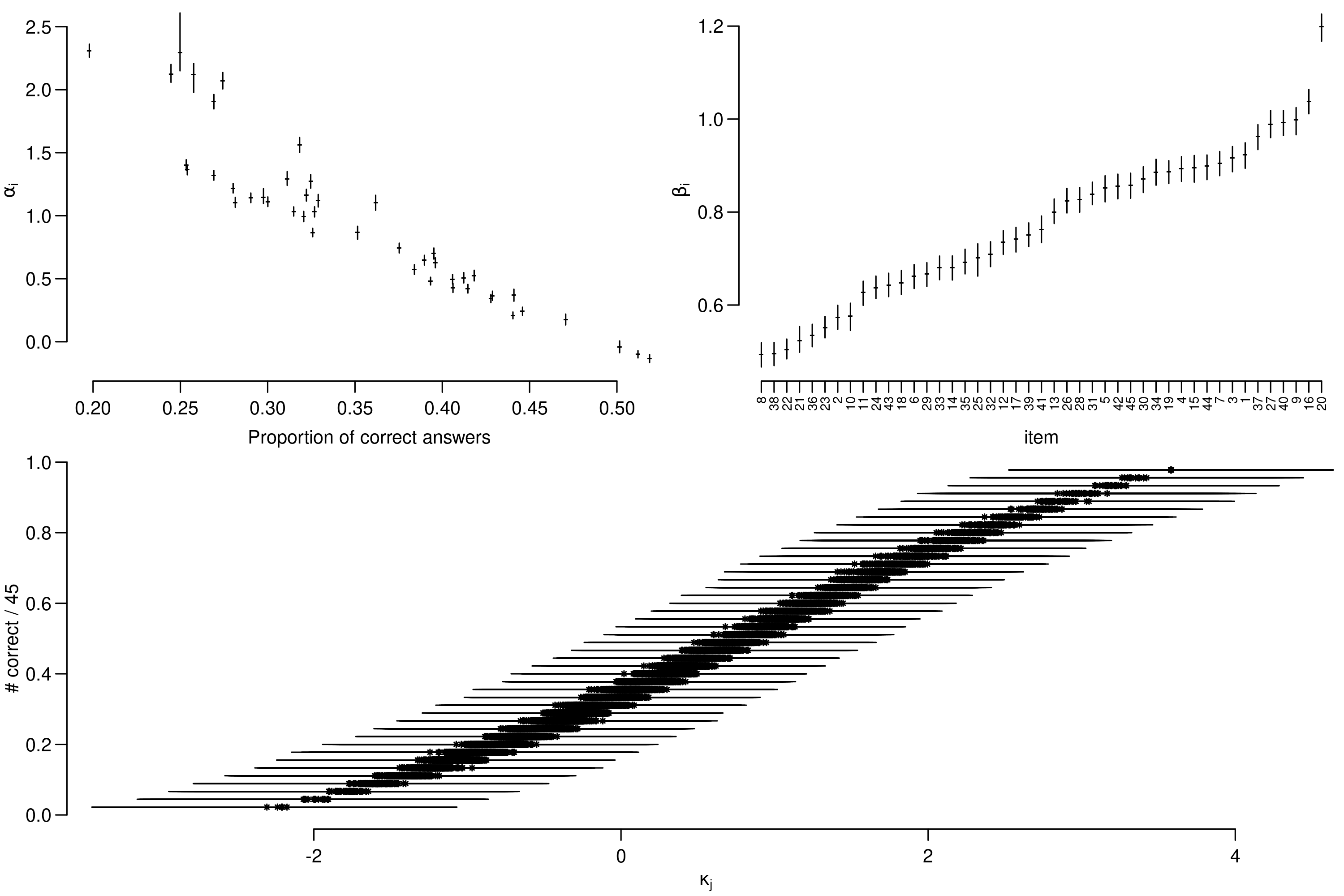}
    \caption{Summary for the model parameters on the ENEM example.}
    \label{irtfig}
\end{figure}

We consider data from the Brazilian national high school exam, the
\textit{Exame Nacional do Ensino Medio} - ENEM. We selected a
Mathematics test which has 45 items, taken by 24956 students of
Curitiba city in 2020. In Figure~\ref{irtfig} we have a summary plots
of the posterior mean and 95\%CI for all the model parameters. The
results were computed in $378$ seconds.

We can see that higher $\alpha_i$ was generally the case for items
with lower proportion of correct answers. For the discrimination
parameter we can see that $\beta_{20}$ is the highest. The students
capabilities are overall proportional to the proportion of items
answered correctly. Overall, students with more correct answers 
have higher $\kappa_j$. The variation in the fitted score
for students with same number of correct answers is due to the
difference in the item parameters. All these students correctly
answered at least one item. The students with all 45 items correctly
answered have the same fit for $\kappa_j$, and the highest mean.

\subsection{Spatial case: a non data-rich case}

In the analysis of data considering the spatial location, the models are
defined so that the size of the random effect accounting for the
spatial correlation, is equal or in the same order as the size of the
data. In disease mapping, a common approach to analyze counts from
areal data is to include two random effects, one structured and one
unstructured, each one with size equal to the number of
observations. In the analysis of point referenced data using the SPDE
approach, the size of the model is usually not the same as the data
size as it is defined on a mesh (see \cite{bakka2018}) with the number of
nodes similar to the number of data locations.  
\\
To illustrate the benefit of the modern INLA formulation for these cases, we
consider an SPDE model with $948$ mesh nodes (see
supplementary material for details of the data generation procedure). We consider the same mesh and
vary the number of data locations, $n$. In Table~\ref{tab1spatial} we
report the number of observations and the computing time considering
the classic and modern INLA approaches. It is clear that even for non data-rich models, the computing time based on
the modern INLA formulation is still low.

\begin{table}[ht]
	\centering
	\begin{tabular}{rrr}
		\hline
		n & Classic INLA (s) & Modern INLA (s) \\ 
		\hline
		$10^2$ & 2.6 & 2.7 \\ 
		$10^3$ & 2.9 & 2.5 \\ 
		$10^4$ & 13.4 & 3.7 \\ 
		\hline
	\end{tabular}
	\caption{Number of data points and the computing time (in seconds)
		considering the classic and modern INLA formulations}
	\label{tab1spatial}
\end{table}

A second point is that one of the goals is to make
predictions at a fine resolution, usually on a fine grid. There are
two ways to compute the posterior marginals at each grid cell in
\textit{INLA}. One way is to include the prediction scenario with the
observations while fitting the model, as the computations are made for
the entire linear predictor. Another way is to fit the model without
the prediction scenario and then draw Monte Carlo samples from the
fitted model and do the computations from these samples. When the
number of predictions is large, the latter was usually suggested (
see \cite{krainski2018advanced}), as the first may cause numerical
instability. This numerical instability stems from the fact that
the classic approach enlarges the model problem by including the
linear predictor in the latent field and this part is near
singular, see Eq. (4) in \citep{rue2017review}. The modern formulation is
thus twofold beneficial, as it avoids the numerical instability and it
reduces the computing time.

We can fit the model considering only the available data, and then
re-fit including the prediction scenario but considering the
previously found mode for the hyperparameters. We consider this
approach for the following evaluation. We evaluated the computing time
considering three different prediction grid sizes for the model on a mesh with $946$
nodes, conditional on $10^4$ observations. In Table~\ref{tab2spatial}
we report the computing time. The reduction in the computing time is
expressive for the finer grid setting. It also scales better for the
new approach when increasing the size of the prediction grid.

\begin{table}[ht]
	\centering
	\begin{tabular}{rrrr}
		\hline
		grid layout & size & Classic INLA (s) & Modern INLA (s)  \\ 
		\hline
		250$\times$150 &  37500 &   5.39 &  2.59 \\ 
		500$\times$300 & 150000 &  17.70 &  4.68 \\ 
		1000$\times$600 & 600000 & 156.48 & 13.69 \\ 
		\hline
	\end{tabular}
	\caption{Number of predictions (grid layout and size) and the
		computing time (in seconds).}
	\label{tab2spatial}
\end{table}

\section{Application to functional magnetic resonance imaging (fMRI)
    models}\label{sec:fmri}

Functional magnetic resonance imaging (fMRI) is a noninvasive
neuro-imaging technique used to localize regions of specific brain
activity during certain tasks (\cite{lindquist2008}). Traditional
volumetric fMRI data consist of a time series of three-dimensional
brain volumes, each of which is composed of thousands of voxels
(equally sized volumetric elements). Volumetric fMRI provides
information on the gray matter (where brain activity occures) as well
as for white matter and cerebral spinal fluid. Mostly, we are
interested in the gray matter and to this end cortical surface fMRI
(cs-fMRI) was proposed as an alternative representation where the
cortical gray matter is represented as a two-dimensional manifold
surface (\cite{fischl2012}).

Cs-fMRI offers dimension reduction and greater neurobiological
significance of distances, amongst many other advantages. Nearby
regions on cs-fMRI tend to exhibit similar behaviour while large
differences can occur between nearby locations in volumetric fMRI. We
can thus use a surface-based spatial statistical model to analyze
cs-fMRI data, proposed by \cite{mejia2020}, defined for $T$ timepoints
and $N$ vertices per hemisphere resulting in data
$\pmb y_{TN\times 1}$ with the latent Gaussian model as follows:
\begin{eqnarray*}
  \pmb y|\pmb\beta, \pmb b, \pmb\theta &\sim& N(\pmb\mu_y, \pmb V),\quad \pmb\mu_y = \sum_{k=0}^K\pmb X_k\pmb\beta_k + \sum_{j=1}^J\pmb Z_j\pmb b_j\\
  \pmb\beta_k &=& \pmb\Psi_k\pmb w_k \quad\text{(SPDE prior on $\pmb\beta_k $, the coefficients of the activation amplitudes $\pmb X_k$)}\\
  \pmb w_k|\pmb\theta &\sim& N(\pmb 0, \pmb Q^{-1}_{\tau_k,\kappa_k})\\
  \pmb b_j &\sim& N(\pmb 0, \delta\pmb I) \quad\text{(Diffuse priors for $\pmb b_j $, the coefficients of the nuisance signals $\pmb Z_j$)}\\
  \pmb\theta &\sim& \pi(\pmb\theta),
\end{eqnarray*}
where we have $K$ task signals and $J$ nuisance signals.

We analyze test-retest motor task fMRI data from a subject in the
Human Connectome Project (HCP)
(\url{https://www.humanconnectome.org/}). Multi-subject analysis can
be done as in \cite{mejia2020} and \cite{spencer2022}, but for
illustration of the modern INLA formulation, we focus on the single
subject case.

The data consists of a 3.5-min fMRI for each subject, consisting of
284 volumes, where each subject performs 5 different motor tasks
interceded with a 3 second visual cue. Each hemisphere of the brain
contained 32492 surface vertices. From these, 5000 are resampled to
use for the analysis. This results in a response data vector $\pmb y$
of size 2 523 624, with an SPDE model (see \cite{lindgren2011},
\cite{bakka2018} and \cite{krainski2018advanced} for details)
defined on a mesh with 8795 triangles. Clearly, the data is much
larger than the size of the model, constituting a data-rich model.
This imbalance in data and model size is inherent to fMRI data and
thus the modern formulation of INLA is essential for performing full
Bayesian inference.

In this illustrative example we consider only the left hemisphere and
4 motor tasks so that we have 4 different activation profiles, each
with an SPDE prior with a common mesh. The inference based on the
modern formulation of INLA was computed in 148 seconds.

\begin{figure}[h]
    \includegraphics[width = 2.8cm]{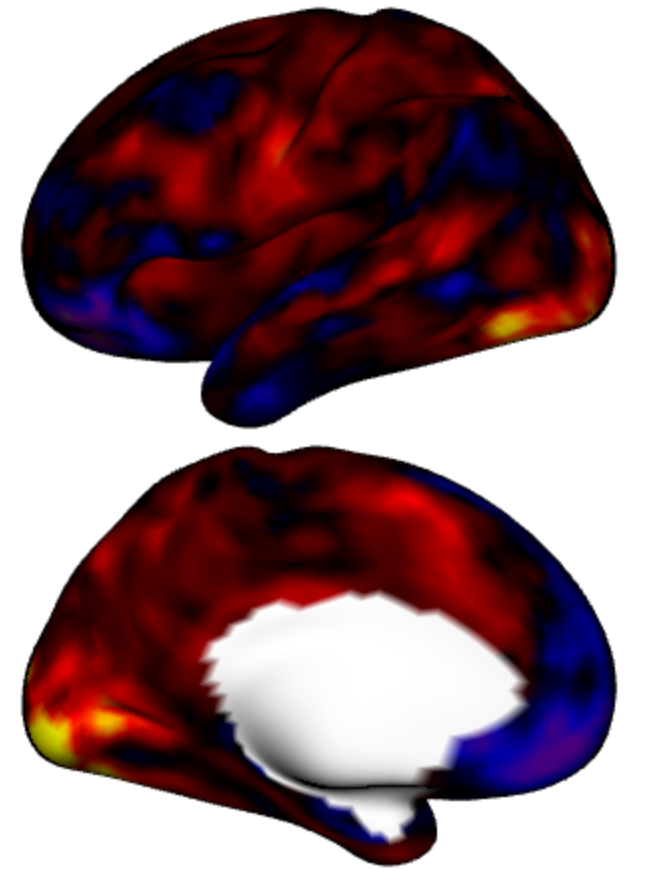}
    \includegraphics[width = 2.8cm]{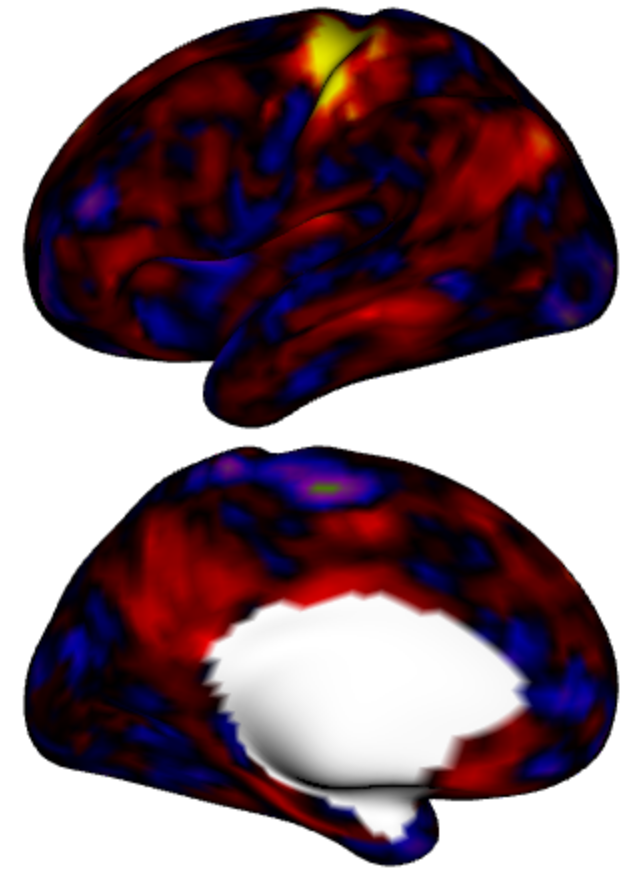}
    \includegraphics[width = 2.8cm]{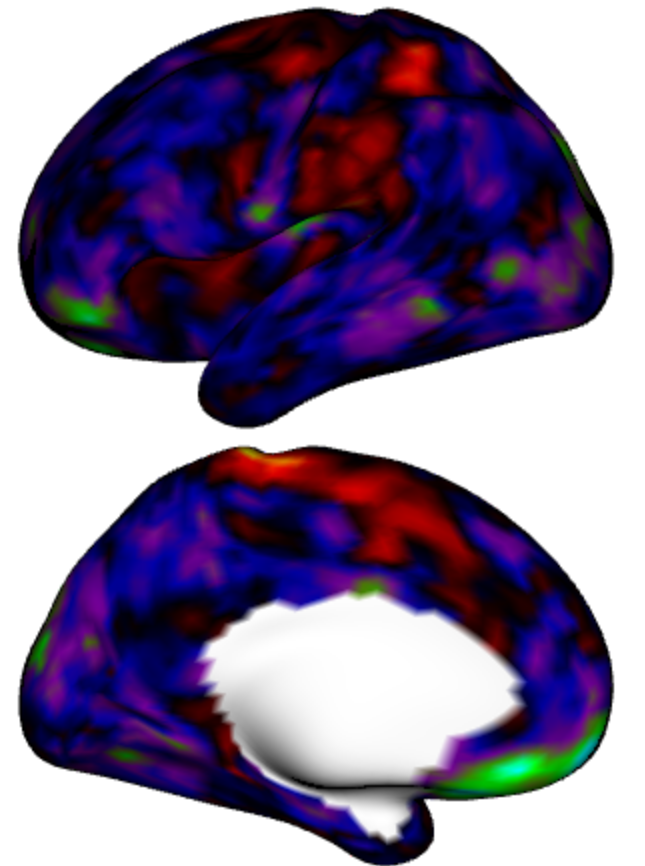}
    \includegraphics[width = 2.8cm]{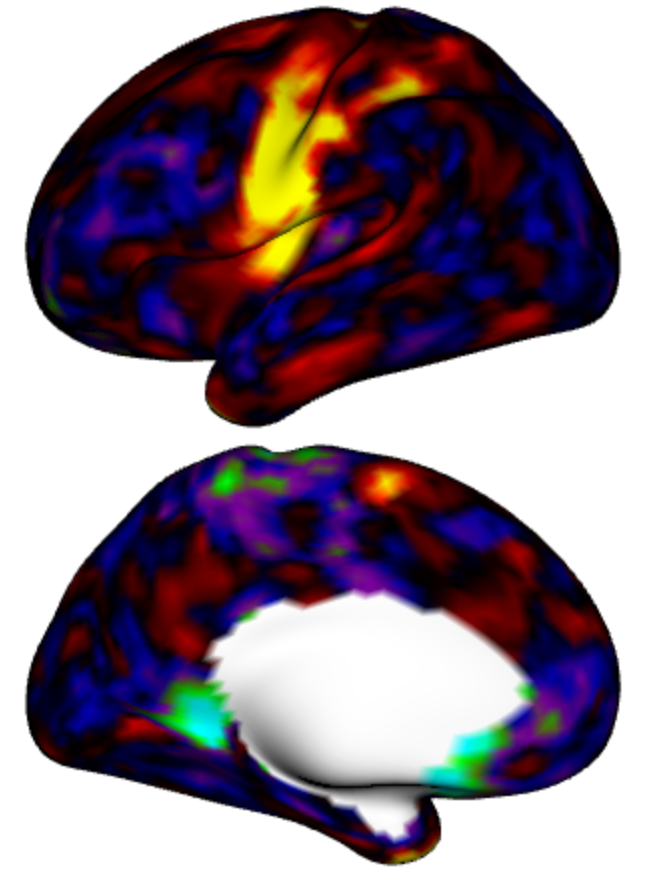}
    \caption{Activation areas for the different tasks in the left
        hemisphere - visual cue, right hand motor, right foot motor,
        tongue motor task (from left to right)}
    \label{fig:fmri}
\end{figure}

\section{Discussion}

We present here a modern formulation of the INLA methodology, based on the reformulation of the latent field by removing the linear predictors from the model. This new framework stems from computational demands set by a certain class of models where the data is large and the model is moderate or small, termed data-rich models. The classic formulation of INLA has paved the way for the many developments within the methodology hitherto, but the need for the modern formulation became clear for the future development and continuing applicability of INLA. Applications of data-rich models using INLA has not been attainable with the classic formulation due to the synthetic enlargement of the latent model size by the linear predictors being included as part of the model components. Now, with the modern formulation, INLA can be used for full Bayesian inference of data-rich models such as spatio-temporal models for fMRI data. \\ \\
The modern formulation, however, became possible only subsequent to the work of \cite{van2021vb}. Since the linear predictors are removed from the model, the Bayesian inference thereof should now be done after the inference of the model is complete. This endeavor is not computationally efficient, unless a Gaussian posterior is assumed for the latent field model, conditional on the hyperparameters. But, the Gaussian assumption might lead to inaccuracies in the estimates which could compound to the linear predictors. A low-rank joint correction to the mean of the latent field, proposed by \cite{van2021vb}, achieves accuracy on par with 'exact' MCMC methods while being computationally efficient with essentially no additional cost. Without this efficient correction, the computational cost associated with the inference of the linear predictors would negate the benefit of removing them from the model. The modern formulation is currently
implemented in R-INLA by the command
\begin{verbatim}
	inla(..., inla.mode = "experimental")
\end{verbatim}
and will be enabled by default in the near future.\\ \\
The advantage of the proposed work is not only computational but it forms the basis of future developments in INLA, for efficient and accurate Bayesian inference of latent Gaussian models. The ideas presented can be used in other ways, like a joint variance and/or skewness correction for the Gaussian assumption, which are currently work in progress within the team. More efficient and accurate hyperparameter marginals are also of current and future interest within the modern framework. For the end user, we believe this work contributes to the toolbox of many
applied statisticians and scientists, who can now perform Bayesian
inference and stable prediction, also for data-rich models, in a computationally efficient and
accurate manner. 

\appendix
\section{Code for the examples}

The code for the examples and application is available at
{\small\url{https://github.com/JanetVN1201/Code_for_papers}}.

\bibliography{BioJ,mybib}

\end{document}